\documentclass[12pt]{iopart}
\usepackage{iopams}  
\usepackage{tipa}

\usepackage{fixmath}
\usepackage{amsfonts}
\usepackage{graphicx}
\usepackage{subfigure}

\def\<{\langle}
\def\>{\rangle}

\begin{document}
\title[Enhanced quantum Zeno effect in the bosonic system]{Landau-Zener transition stabilized by the enhanced quantum Zeno effect in the bosonic system}
\author{Kai Wen$^1$, Tim Byrnes$^{2}$, and Yoshihisa Yamamoto$^{1,2}$}
\address{$^1$ Edward L. Ginzton Laboratory, Stanford University, Stanford, California 94305, USA\\
$^2$ National Institute of Informatics, 2-1-2 Hitotsubashi,
Chiyoda-ku, Tokyo, 101-843, Japan}
\date{\today}
\ead{kaiwen@stanford.edu}
\begin{abstract}
We study the Landau-Zener transition with the quantum Zeno effect in an open dissipative system populated by a large number of bosons. Given the quantum Zeno effect is strong enough, both discrete and continuous quantum Zeno measurements are found to stabilize the Landau-Zener transition. Both the $\sigma^x$-type longitudinal relaxation and $\sigma^z$-type transverse relaxation in the bosonic system are analyzed as a model of continuous quantum Zeno measurements. While both of them improve the signal-to-noise ratio in terms of the ground state population, the $\sigma^x$-type relaxation can further boost measurement sensitivity and thus lead to a polynomial speedup with the number of bosons in the system. For a system that contains a large number of bosons such as in a Bose-Einstein condensate with more than $10^4$ bosons, this equates to several orders of magnitude speedup.
\end{abstract}


\maketitle

\section{Introduction}

In the quantum Zeno effect, if a quantum system is frequently projected in a certain measurement basis including the initial state, the system is pinned on this initial state\cite{zeno}. Moreover, if the sequence of measurements is performed along a specific path, the system will evolve along this arbitrarily chosen path. This effect is called the inverse quantum Zeno effect\cite{inv-zeno1, inv-zeno2} and has been proposed to manipulate quantum systems by Kitano\cite{Kitano}.  The measurement basis of the sequence typically follows the diagonal basis of the temporal Hamiltonian of an adiabatic evolution. The goal of the control sequence is to use the inverse quantum Zeno effect to keep the system in the ground state which is one of the eigenstates of the temporal Hamiltonian. In this sense, the inverse quantum Zeno effect can also solve the problems targeted by adiabatic quantum computation\cite{aqc}. However, the physics behind the measurement-based evolution is a dephasing and dissipation with respect to the reservoir, while adiabatic evolution is completely coherent. Thus the two schemes yield different time complexity\cite{Kitano}. 

More recent developments have revealed that quantum computation by engineered dissipation and decoherence is a promising alternative to coherent quantum computation \cite{open_qc1, open_qc2, open_qc3}. It is also well-known that bosonic systems are able to enhance the dissipation process. One example is the property of final state stimulation, whereby in a dissipative process, if $N$ bosons are present in the final state, then the cooling rate is enhanced by a factor of $N+1$. A classical computational scheme that takes advantage of bosonic final state stimulation was proposed in Ref. \cite{bec}. It was found there that the time for solving to a typical Ising model type problem could be accelerated by a factor of $1/N$, where $N$ is the number of bosons per Ising model site. 

In this paper we study the performance of quantum computation by the quantum Zeno effect in an open dissipative system for a many-boson system. Note that henceforth we do not discriminate between the quantum Zeno effect and the inverse quantum Zeno effect. We focus on the scheme involving measurements along the path of the Hamiltonian for a Landau-Zener transition\cite{LZ, LZ2}, the simplest example of an adiabatic quantum computation. We show that the Landau-Zener transition can be stabilized by a series of strong enough quantum Zeno measurements. More importantly, the quantum Zeno effect is enhanced significantly in the presence of many bosons and thus can accelerate the scheme of the Landau-Zener transition by improving the measurement sensitivity.

\section{Discrete model}

First, we show that the Landau-Zener transition can be realized by the discrete quantum Zeno effect by a series of projective measurements. In the bosonic system, the Hamiltonian for the Landau-Zener transition is written as
\begin{eqnarray}
\hat{H} = -\epsilon(t) \hat{J}^z - \Delta\hat{J}^x, \label{eq:H_lz}
\end{eqnarray}
where $-T/2 \leq t \leq T/2$, and $\epsilon(t) = 2\epsilon_0 t/T = v t$.  $\hat{J}^z =  ( \hat{a}^\dag \hat{a} - \hat{b}^\dag \hat{b} )/2$ and $\hat{J}^x =  ( \hat{a}^\dag \hat{b} + \hat{b}^\dag \hat{a} )/2$, where $\hat{a}$ and $\hat{b}$ are the bosonic annihilation operators for the two energy levels, for example, up-spin and down-spin states, respectively. Suppose that $\epsilon_0 \gg \Delta > 0$, and the system of $N$ bosons is initialized in the initial ground state $|0\>_a |N\>_b$. By evolving the Hamiltonian slower than the adiabatic limit, the system is pinned in the ground state and reaches the state $|N\>_a |0\>_b$ at $t=T/2$.

The same goal can be achieved by a discrete quantum Zeno effect, consisting of a series of projective measurements along the diagonal basis of the time-dependent Hamiltonian given in Eq. (\ref{eq:H_lz}). We can diagonalize Eq. (\ref{eq:H_lz}) as
\begin{eqnarray}
\hat{H} &=& -\frac{\sqrt{\epsilon(t)^2+\Delta^2}}{2} ( \hat{P}^\dag(t)
\hat{P}(t) - \hat{Q}^\dag(t) \hat{Q}(t)), \label{eq:Hdiag}
\end{eqnarray}
where the time-dependent annihilation operators are defined as $\hat{P} = U \hat{b} + V \hat{a}, 
\hat{Q} = -V \hat{b} + U \hat{a}$, and $|U|^2 = [1-\epsilon(t)/\sqrt{\epsilon(t)^2+\Delta^2}]/2, |V|^2 =[1+\epsilon(t) / \sqrt{\epsilon(t)^2+\Delta^2}]/2$. Note that $|U|^2 + |V|^2 = 1$. We assume the system is initialized in the state $|N\>_P|0\>_Q$.

Now suppose we divide the evolution time from $-T/2$ to $T/2$ into $n$ time intervals. At $t_k=-T/2 + k \delta t$, where $\delta t = T/n$ and $k = 1, \cdots, n$, we perform the measurement on each boson in the diagonal basis of the temporal Hamiltonian. Suppose at $t=t_{k}$ the system is projected to $|N\>_{P_k}|0\>_{Q_k}$. The probability that the system is projected to $|N\>_{P_{k+1}}|0\>_{Q_{k+1}}$ at $t=t_{k+1}$ is given by
\begin{eqnarray}
\textrm{Prob}_{k+1} = (U_{k+1} U_{k} + V_{k+1} V_{k})^N. \label{eq:pk}
\end{eqnarray}

We define $\delta \epsilon = v \delta t = 2 \epsilon_0/n$ and the dimensionless parameter $\delta r_{k+1} = \delta \epsilon / \sqrt{\epsilon(t_{k})^2+\Delta^2} \leq \delta\epsilon/\Delta$ which can be arbitrary small when $n$ is large enough (i.e. the frequency of measurement is high enough). Then we can approximate Eq. (\ref{eq:pk}) as 
\begin{eqnarray}
\textrm{Prob}_{k+1} = \left[ 1 - \frac{\Delta^2}{8(\epsilon(t_{k})^2+\Delta^2)}\Delta r_{k+1}^2 + o(\Delta r_{k+1}^2)\right]^N \geq \left[1-\frac{\delta \epsilon^2}{8\Delta^2}\right]^N.
\end{eqnarray}

The probability of the state being in the ground state at $t=t_n=T/2$ is then bounded by
\begin{eqnarray}
\textrm{Prob}(|N\>_{P_n}|0\>_{Q_n}) \geq \prod_{k=1}^n \textrm{Prob}_k  \geq \left[1 - \frac{\epsilon_0^2}{2\Delta^2}\times\frac{1}{n^2}\right]^{Nn}
\end{eqnarray}
The RHS goes to 1 when $n$ goes to infinity, therefore 
\begin{eqnarray}
\lim_{n \rightarrow \infty} \textrm{Prob}(|N\>_{P_n}|0\>_{Q_n}) = 1.
\end{eqnarray}
This result shows that the quantum Zeno effect is also capable of stabilizing the Landau-Zener transition by discrete projective measurements. The success of the quantum Zeno effect only depends on the measurement frequency $n$ and the ratio $\epsilon_0/\Delta$ . Stabilization of the ground state can always be achieved by choosing $n$ sufficiently large, as long as there is a A finitenon-zero minimum energy gap $\Delta$.

\section{Continuous model}

Although such projective measurements can in principle stabilize the Landau-Zener transition, practically such a sequence is very difficult to implement. The reason for this is that knowledge of all the instantaneous eigenstates along the adiabatic sweep is required in order to carry out such a procedure.  This is in general more difficult than finding the solution state, which is just one of the eigenstates.

To utilize the quantum Zeno effect in a practical quantum computation setting, let us now include a system-reservoir coupling.  Suppose that we continuously evolve the system Hamiltonian and open the system to the environment. The decoherence induced by the system-reservoir coupling is mainly in the diagonal basis of the system Hamiltonian, assuming that the system-reservoir coupling is much faster than the evolution of the system Hamiltonian. The coupling to the reservoir can be described as a continuous measurement in the diagonal basis of the instantaneous Hamiltonian implementing the quantum Zeno measurements. This passive measurement is much more practical since it is achieved without knowing the actual eigenstates of the Hamiltonian.

By evolving the system Hamiltonian according to the scheme of quantum adiabatic computation, the continuous measurement induced by the system-reservoir coupling can realize the quantum Zeno effect and pin the system in the ground state. This evolves the system to the final ground state, which is the solution of the computational problem. 

The decoherence due to the system-reservoir coupling is usually decomposed into two types: a $\sigma^x$-type longitudinal relaxation and a $\sigma^z$-type transverse relaxation. The total master equation for both types of decoherence at zero temperature is written as
\begin{eqnarray}
\frac{d\rho}{dt} = -i [\hat{H}, \rho] + \frac{\Gamma_x}{2} \mathcal{D}\left[\tilde{J}^-\right] \rho + \frac{\Gamma_z}{2} \mathcal{D}\left[\tilde{J}^z\right] \rho ,\label{eq:master-zeno}
\end{eqnarray}
where $\hat{H}$ is given in Eq. (\ref{eq:Hdiag}), and $\mathcal{D}[\hat{c}]\rho = - \hat{c}^\dag \hat{c}\rho - \rho \hat{c}^\dag \hat{c} + 2 \hat{c}\rho\hat{c}^\dag$ is the Linblad operator. $\tilde{J}^- = \hat{P}^\dag\hat{Q}$ is the operator for the $\sigma^x$-type longitudinal relaxation, and $\Gamma_x$ is the associated relaxation rate assumed to be constant. $\tilde{J}^z = (\hat{P}^\dag\hat{P} - \hat{Q}^\dag\hat{Q})/2$ is the operator for the $\sigma^z$-type transverse relaxation, and $\Gamma_z$ is the associated relaxation rate also assumed to be constant. We consider the system to be at zero temperature for simplicity.

First, we solve the master Eq. (\ref{eq:master-zeno}) only with the $\sigma^x$-type relaxation, by following the method in Ref. \cite{Agarwal}. The average of any operator $\hat{A}$ on the system is given by 
\begin{eqnarray}
\frac{d}{d t} \<\hat{A}\> &=& i \sqrt{\epsilon(t)^2+\Delta^2} \<[\tilde{J}^z, A]\> + \<\frac{\partial}{\partial t}\hat{A}\> \nonumber \\
&&- \frac{\Gamma_x}{2} \<\tilde{J}^+ \tilde{J}^- \hat{A} - 2 \tilde{J}^+ \hat{A} \tilde{J}^- + \hat{A} \tilde{J}^+ \tilde{J}^-\>.
\end{eqnarray}
We then derive the equation for $\tilde{J}^z_i$ on the $i$-th boson as 
\begin{eqnarray}
\frac{d}{d t} \<\tilde{J}^z_i\> &=& - \kappa(t) ( \<\tilde{J}^+_i\> + \<\tilde{J}^-_i\>) - \frac{\Gamma_x}{2} (\tilde{J}^+_i \tilde{J}^-_i \tilde{J}^z_i - 2 \tilde{J}^+_i \tilde{J}^z_i \tilde{J}^-_i + \tilde{J}^z_i \tilde{J}^+_i \tilde{J}^-_i ) \nonumber \\
&& - \frac{\Gamma_x}{2} \sum_{j \neq i}( \tilde{J}^+_j \tilde{J}^-_i \tilde{J}^z_i - 2 \tilde{J}^+_j \tilde{J}^z_i \tilde{J}^-_i + \tilde{J}^z_i \tilde{J}^+_j \tilde{J}^-_i \nonumber \\
&& + \tilde{J}^+_i \tilde{J}^-_j \tilde{J}^z_i - 2 \tilde{J}^+_i \tilde{J}^z_i \tilde{J}^-_j + \tilde{J}^z_i \tilde{J}^+_i \tilde{J}^-_j).
\end{eqnarray}
The first bracket on the RHS comes from $\<\frac{\partial}{\partial t}\tilde{J}^z_i\>$, where $\kappa(t) = (\dot{V} U - \dot{U} V) = \frac{\dot{V}^2 U^2 - \dot{U}^2 V^2}{2 U V} = \frac{\dot{\epsilon}(t) \Delta}{2(\epsilon(t)^2 + \Delta^2)}$ characterizes the strength of the diadiabatic excitation. The last two brackets on the RHS come from the $\sigma^x$-type relaxation. By using $\tilde{J}^z_i = -\frac{1}{2} + \tilde{J}^+_i \tilde{J}^-_i$, the second bracket is evaluated as $-\Gamma_x(\frac{1}{2}+\<\tilde{J}^z_i\>)$. The third bracket is then transformed into $-\frac{\Gamma_x}{2}\sum_{j \neq i} (\<\tilde{\bold{J}}_i \cdot  \tilde{\bold{J}}_j\> - 2\<\tilde{J}^z_i \tilde{J}^z_j\>)$, where $\<\tilde{\bold{J}}_i \cdot  \tilde{\bold{J}}_j\> = \<\tilde{J}^x_i \tilde{J}^x_j + \tilde{J}^y_i \tilde{J}^y_j + \tilde{J}^z_i \tilde{J}^z_j\>$ can be easily proved to be invariant during the evolution. With the initial condition that $|\psi(t=-T/2)\> = |N\>_P|0\>_Q$, we find that $\<\tilde{\bold{J}}_i \cdot  \tilde{\bold{J}}_j\> \equiv 1/4$, for any $j \neq i$. Combining equations for all $\tilde{J}^z_i$, we obtain 
\begin{eqnarray}
\frac{d}{d t} \<\tilde{J}^z\> &=& - \kappa(t) ( \<\tilde{J}^+\> + \<\tilde{J}^-\> ) - \Gamma_x (\frac{1}{2}N+\<\tilde{J}^z\>)  \nonumber \\
&& - \frac{\Gamma_x}{2}[\frac{1}{2}N(N-1) - 2 \sum_{i}\sum_{j \neq i}\<\tilde{J}^z_i \tilde{J}^z_j\>].
\end{eqnarray}
Using $\tilde{J}^z_i =  - \frac{1}{2} + \tilde{J}^+_i \tilde{J}^-_j$ again, $\sum_{i}\sum_{j \neq i}\<\tilde{J}^z_i \tilde{J}^z_j\> = -N(N-1)/4 - (N-1)\<\tilde{J}^z_i\> + N(N-1)\<\tilde{J}^+_i \tilde{J}^-_i \tilde{J}^+_j \tilde{J}^-_j\>$, for any $i \neq j$. Finally, we use the first-order Hartree approximation\cite{Agarwal}, i.e.,  $\<\tilde{J}^+_i \tilde{J}^-_i \tilde{J}^+_j \tilde{J}^-_j\> \approx \<\tilde{J}^+_i \tilde{J}^-_i\> \<\tilde{J}^+_j \tilde{J}^-_j\> = (1/2 + \<\tilde{J}_z\>/N)^2$, to obtain
\begin{eqnarray}
\frac{d}{d t} \<\tilde{J}^z\> &=& - \kappa(t) \left( \<\tilde{J}^+\> +
\<\tilde{J}^-\> \right) - \Gamma_x \left(\frac{1}{2}N+
\<\tilde{J}^z\>\right)
\nonumber \\
&&- \Gamma_x (N-1)\left(\frac{N}{4} -
\frac{\<\tilde{J}^z\>^2}{N}\right), \label{eq:sigmax_Jz}
\end{eqnarray}

A similar method is used to derive the equations for $\<J^\pm\>$ with $\sigma^x$-type relaxation. We first obtain the equations for $\<J^\pm_i\>$ on the $i$-th boson as 
\begin{eqnarray}
\frac{d}{dt}\<\tilde{J}^\pm_i\> &=& \pm i \sqrt{\epsilon(t)^2 + \Delta^2}\<\tilde{J}^\pm\> + 2\kappa(t)\<\tilde{J}^z_i\> \nonumber \\
&&- \frac{\Gamma_x}{2}\<\tilde{J}^\pm_i\> - \Gamma_x \sum_{j \neq i} \<\tilde{J}^+_i \tilde{J}^-_i \tilde{J}^\pm_j\>.
\end{eqnarray}
Then we employ a similar approximation, namely, $\<\tilde{J}^+_i \tilde{J}^-_i \tilde{J}^+_j \tilde{J}^-_j\> \approx \<\tilde{J}^+_i \tilde{J}^-_i\> \<\tilde{J}^+_j \tilde{J}^-_j\>$ and $\<\tilde{J}^+_i \tilde{J}^-_i \tilde{J}^\pm_j\> \approx \<\tilde{J}^+_i \tilde{J}^-_i\> \<\tilde{J}^\pm_j\>$ for any $i \neq j$. The result is
\begin{eqnarray}
\frac{d}{d t}
\<\tilde{J}^\pm\> &=& \pm i \sqrt{\epsilon(t)^2+\Delta^2}
\<\tilde{J}^\pm\> + 2 \kappa(t) \<\tilde{J}^z\> -
\frac{\Gamma_x}{2} \<\tilde{J}^\pm\> \nonumber \\
&&- \Gamma_x
(N-1)\<\tilde{J}^\pm\>\left(\frac{1}{2}+\frac{\<\tilde{J}^z\>}{N}\right)
\label{eq:sigmax_Jpm}.
\end{eqnarray}

Second, we solve the master Eq. (\ref{eq:master-zeno}) only for $\sigma^z$-type
transverse relaxation. Similar to the first case, we use
\begin{eqnarray}
\frac{d}{d t} \<\hat{A}\> &=&  i \sqrt{\epsilon(t)^2+\Delta^2}
\left\<[\tilde{J}_z, \hat{A}]\right\> + \left\<\frac{\partial}{\partial t}\hat{A}\right\> \nonumber \\
&&-
\frac{\Gamma_z}{2} \<\tilde{J}^z \tilde{J}^z \hat{A}  + \hat{A} \tilde{J}^z \tilde{J}^z - 
2 \tilde{J}^z \hat{A} \tilde{J}^z\>.
\end{eqnarray}
We can easily derive the equations for the angular
momentum for a single boson and extend the results to those on the collective operators as
\begin{eqnarray}
\frac{d}{d t} \<\tilde{J}^z\> &=& - \kappa(t) \left( \<\tilde{J}^+\>
+ \<\tilde{J}^-\> \right), \label{eq:sigmaz_Jz}\\
\frac{d}{d t} \<\tilde{J}^\pm\> &=& \pm i
\sqrt{\epsilon(t)^2+\Delta^2} \<\tilde{J}^\pm\> + 2 \kappa(t)
\<\tilde{J}^z\> - \frac{\Gamma_z}{4} \<\tilde{J}^\pm\> \label{eq:sigmaz_Jpm}.
\end{eqnarray}

Equations (\ref{eq:sigmax_Jz}), (\ref{eq:sigmax_Jpm}), (\ref{eq:sigmaz_Jz}) and (\ref{eq:sigmaz_Jpm}) show that if the relaxation rates $\Gamma_z$ and $\Gamma_x$ are large enough, $\<\tilde{J}^\pm\>$ is kept as $0$. Therefore the system is almost completely in the ground state at all times and $\<\tilde{J}^z\>$ is close to $-N/2$, even if the evolution is faster than the adiabatic limit. This shows that the Landau-Zener transition can be stabilized by the continuous quantum Zeno effect induced by a strong enough system-reservoir coupling. Note that a finite minimum bandgap $\Delta$ is also necessary to make the Landau-Zener transition successful with the quantum Zeno effect. Without this condition, $\kappa(t)$ diverges and $\<\tilde{J}^\pm\>$ cannot be kept to $0$ under any condition.

\begin{figure}[h]
\begin{center}
\includegraphics[width=8cm]{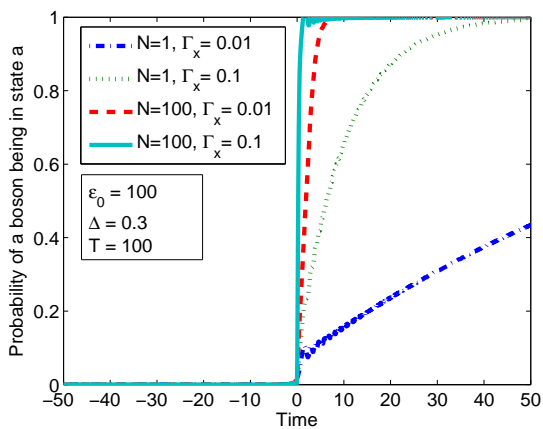}
\caption{(Color online) The probability of an individual boson being in state $a$ during the evolution with a $\sigma^x$-type longitudinal relaxation for various $N$ and $\Gamma_x$.}\label{fig:zeno-sigmax-lz}
\end{center}
\end{figure}

\begin{figure}[h]
\begin{center}
\includegraphics[width=8cm]{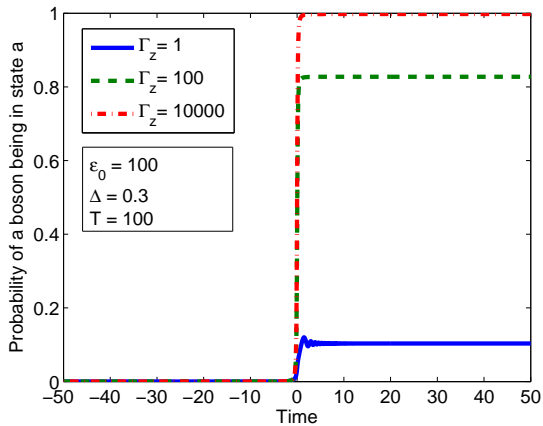}
\caption{(Color online) The probability of an individual boson being in state $a$ during the evolution with a $\sigma^z$-type transverse relaxation for various $\Gamma_z$.}\label{fig:zeno-sigmaz-lz}
\end{center}
\end{figure}

Figures \ref{fig:zeno-sigmax-lz} and \ref{fig:zeno-sigmaz-lz} demonstrate the time-dependent evolutions of the Landau-Zener transition with a $\sigma^x$-type longitudinal relaxation and a $\sigma^z$-type transverse relaxation. In both cases, the initial condition is $|\psi(t=-T/2)\> = |N\>_P|0\>_Q$. Both results demonstrate that by increasing the relaxation strength the quantum Zeno effect can successfully stabilize the Landau-Zener transition. However, significant differences between the two cases exist. First, in Fig. \ref{fig:zeno-sigmax-lz}, the presence of many bosons enhances the effective relaxation strength as can be seen by examining the dependence with $N$ in Eq. (\ref{eq:sigmax_Jz}) and (\ref{eq:sigmax_Jpm}). Even if $\Gamma_x$ is very small, a large $N$ makes the system evolution approach to the ``quantum Zeno effect limit''. On the other hand, for the case of $\sigma^z$-type relaxation in Eqs. (\ref{eq:sigmaz_Jz}) and (\ref{eq:sigmaz_Jpm}) , the transverse relaxation rates are independent of $N$. Thus many bosons will not enhance the final probability of the system being in the ground state. 

Second, if the effective relaxation strength is not large enough, the $\sigma^x$-type relaxation can still drive the system back to the ground state by dissipation, for example, by emitting phonons to the reservoir. The relaxation is also enhanced by larger $N$, by comparing the dashed line and the dashed-dotted line in Fig. \ref{fig:zeno-sigmax-lz}. Cases where the system is significantly excited at the anti-crossing point ($t=0$) but relaxes to the ground state after the evolution finishes may be called the ``bosonic final state stimulation limit''. The case of Fig. \ref{fig:zeno-sigmaz-lz} does not have such effect since it only causes dephasing to the system. Summarizing the two points, the $\sigma^x$-type relaxation is advantageous over the $\sigma^z$-type relaxation. With many bosons, both cases can improve the measurement signal-to-noise ratio, but the $\sigma^x$-type relaxation will further boost the measurement sensitivity.

\section{Success probability}

We now turn to calculate the success probability of the Landau-Zener transition with continuous quantum Zeno measurements. As we have shown that the $\sigma^x$-type relaxation is advantageous, we will focus on this type of relaxation to demonstrate the speedup with many bosons.

In the Landau-Zener transition, the answer state is the up-spin state $a$. In tge many-boson system, the algorithm outputs state $a$ if more than half of the bosons are measured to be in state $a$. Given that $p$ is the probability of an individual boson being in state $a$ and the measurement outcome of each boson is either 0 or 1, (state $b$ or $a$ respectively), for large $N$, the ensemble average of the measurement outcome of each boson can be approximated by a Gaussian distribution with mean value $\mu = Np$ and variance $\sigma^2 = Np(1-p)$. As the cut-off is $0.5$, the probability that the algorithm fails is $P_e = [1-\textrm{erf}(\frac{p-0.5}{\sigma\sqrt{2}})]/2$, and the success probability is $1 - P_e$. Therefore, given $P_e < 10^{-12}$, a typical tolerable error rate in communications, we can find the requirement for $p$ and thus simulate the minimum evolution time to achieve this requirement.  Signal-to-noise ratio can be improved by increasing $N$, in the sense that the lower bound for $p$ is decreased. If $N$ is large enough, the improvement of signal-to-noise ratio is less significant. Nevertheless, $\sigma^x$-type relaxation can further boost $p$ (i.e. the measurement sensitivity) by adding more bosons.

\begin{figure}[t]
\begin{center}
\includegraphics[width=8cm]{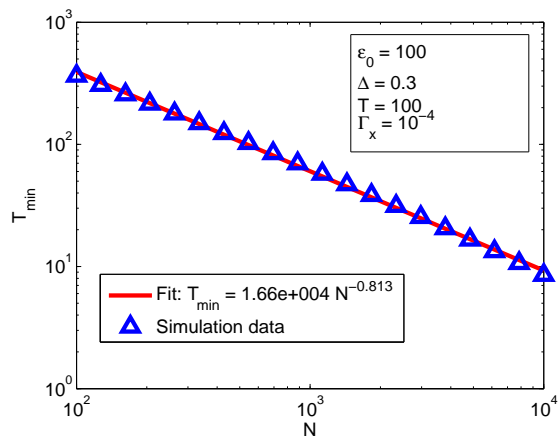}
\caption{(Color online) Simulation of minimum time $T_{\min}$ to reach $P_e < 10^{-12}$ with various $N$.}\label{fig:zeno-lz-minT}
\end{center}
\end{figure}

Fig. \ref{fig:zeno-lz-minT} illustrates the speedup of the Landau-Zener transition with increasing number of bosons in the system. At large $N$, the relation approaches a straight line on the log-log plot. The fitted result gives a polynomial speedup of around $N^{-0.81}$ under the condition that $\sigma^x$-type relaxation plays a significant role at zero temperature. Note that in order to achieve $P_e < 10^{-12}$, the probability of a boson being in state $a$ does not need to be close to 1. Therefore, the algorithm always works in the bosonic final state stimulation region where the system relaxes to the ground state via actual dissipation, for example, by emitting phonons. Although it is only a polynomial speedup, the speedup can be in practice quite large for a large number of bosons. For example, in proposals where Bose-Einstein condensates are used as the effective qubits, there can be more than $10^4$ bosons.

\section{Conclusion}

We have demonstrated that the Landau-Zener transition can be stabilized by both the discrete and the continuous quantum Zeno effect. In the continuous model, two types of relaxation, the $\sigma^x$-type longitudinal relaxation and the $\sigma^z$-type transverse relaxation were analyzed.  The reason for the stabilization of the ground state is that the coupling to a reservoir can give information relating to whether the system has been excited above the ground state.  For example, in the case of strong $\sigma^x$-type relaxation, a negative measurement result, i.e. a lack of excitations in the reservoir (e.g. phonons), is a signal that the system is not being excited. On the other hand, for $\sigma^z$-type relaxation, the measurement readout directly gives information of what state the system is in, since it is an effective measurement in the diagonal basis of the temporal Hamiltonian. In contrast to previous works on adiabatic quantum computation in open dissipative systems \cite{open_aqc1, open_aqc2, open_aqc3, open_aqc4} that have been concerned with the robustness and enhancement of the evolution, our work emphasizes the relaxation which can play a more significant role than the coherent evolution.

By introducing the many-boson system quantum Zeno effect with $\sigma^x$-type relaxation, we have found that this can significantly enhance both the signal-to-noise ratio and measurement sensitivity.  The enhancement is found to be a polynomial speedup of the number of bosons in the system. By use of a large number of bosons, this introduces the possibility of further accelerating quantum computation beyond the standard qubit case.

The authors wish to thank N. Cody Jones for discussions. This research is supported by Navy/SPAWAR Grant N66001-09-1-2024, MEXT, NICT, JSPS through its FIRST Program, and the Special Coordination Funds for Promoting Science and Technology of University of Tokyo, NICT and MEXT.

\end{document}